\documentclass[aps,prapplied,reprint,groupedaddress]{revtex4-2}

\usepackage{graphicx}
\usepackage{subfigure}
\usepackage{float}
\usepackage{dcolumn}
\usepackage{makecell}
\usepackage{bm}
\usepackage{upgreek}
\usepackage[colorlinks=true,linkcolor=blue,citecolor=blue,urlcolor=blue]{hyperref}

\usepackage[abs]{overpic}
\usepackage{siunitx}

\usepackage{xpatch}

\usepackage{pifont}

\begin{document}


\title{Faraday laser pumped cesium beam clock}



\author{Hangbo Shi$^{1}$}
\author{Xiaomin Qin$^{1}$}
\author{Haijun Chen$^{2}$}
\author{Yufei Yan$^{2}$}
\author{Ziqi Lu$^{1}$}
\author{Zhiyang Wang$^{1}$}
\author{Zijie Liu$^{1}$}
\author{Xiaolei Guan$^{1}$}
\author{Qiang Wei$^{3}$}
\author{Tiantian Shi$^{4}$}
\email{tts@pku.edu.cn}
\author{Jingbiao Chen$^{1,5}$}
\affiliation{$^{1}$ State Key Laboratory of Advanced Optical Communication Systems and Networks, Institute of Quantum Electronics, School of Electronics, Peking University, Beijing 100871, China}
\affiliation{$^{2}$ National Key Laboratory of Science and Technology on Vacuum Electronics, Beijing Vacuum Electronics Research Institute, Beijing 100871, China}
\affiliation{$^{3}$ Chengdu Spaceon Electronics Co., Ltd., Chengdu 611731, China}
\affiliation{$^{4}$ National Key Laboratory of Advanced Micro and Nano Manufacture Technology, School of Integrated Circuits, Peking University, Beijing 100871, China}
\affiliation{$^{5}$ Hefei National Laboratory, Hefei 230088, China}

\date{\today}

\begin{abstract}
We realize a high-performance compact optically pumped cesium beam atomic clock utilizing Faraday laser simultaneously as pumping and detection lasers. The Faraday laser, frequency stabilized by modulation transfer spectroscopy (MTS) technique, has narrow linewidth and superior frequency stability. Measured by optical heterodyne method between two identical systems, the linewidth of the Faraday laser is 2.5 kHz after MTS locking, and the fractional frequency stability of the Faraday laser is  optimized to $1.8\times{10}^{-12}/\sqrt{\tau}$. Based on this high-performance Faraday laser, the cesium beam clock realizes a signal-to-noise ratio (SNR) in 1 Hz bandwidth  of $39600$ when the cesium oven temperature is 130°C. Frequency-compared with Hydrogen maser, the fractional frequency stability of the Faraday laser pumped cesium beam clock can reach $1.3\times{10}^{-12}/\sqrt{\tau}$ and decreases to $1.4\times{10}^{-14}$ at 10000 s with the cesium oven temperature set at 110°C.
This Faraday laser pumped cesium beam clock demonstrates its excellent performance, and its great potential in the fields of timekeeping, navigation, and communication. Meanwhile, the Faraday laser, as a high-performance optical frequency standard, can also contribute to the development of other applications in quantum metrology, precision measurement, and atomic physics.

\end{abstract}

\maketitle


\section{Introduction}

Since being constructed\cite{essen}, cesium beam clocks have been widely used in the field of frequency and time, promoting the development of timekeeping, high-speed communication, navigation, and fundamental research\cite{Vanier_2005,Maleki_2005}. There are two main types of cesium beam clocks, magnetic state selecting cesium beam clock\cite{essen1957caesium,1457534,Audoin,510} and optically pumped cesium beam clock\cite{GPS_3,2002,Galileo_2003,Thales_2006,Galileo, shanghaosen,chenhaijun,chenxuzong,OSA-3300}. Compared with magnetic state selecting cesium beam clocks, optically pumped cesium beam clocks have a significantly higher atomic utilization, leading to better frequency stability and longer service life. The optically pumped cesium beam clocks is gradually showing a dominant trend.

The frequency stability of optically pumped cesium beam clocks can be further optimized mainly by reducing the shot noise of cesium atoms and the frequency noise of the laser. Increasing the number of interacting atoms can effectively reduce the shot noise of cesium atoms, which can be realized by increasing the atomic beam flux. In practice, the atomic beam flux is chosen to an appropriate finite value, taking into account the lifetime of the cesium clock.  When the atomic beam flux is constant, the frequency stability of optically pumped cesium atomic clocks can be greatly improved by adopting a laser with narrow linewidth, low frequency noise, and high power stability, which can effectively improve the signal-to-noise ratio (SNR) of Ramsey fringes of the optically pumped cesium beam clock\cite{278532}. At the same time, miniaturization and transportability also need to be considered. The volume of the cesium beam clock is mainly limited by the cesium beam tube, but the shortening of the free drift region will reduce the Ramsey evolution time, which will widen the linewidth of the Ramsey fringes to a certain extent and lead to the the deterioration in frequency stability of cesium beam clock. Thus, the optimization of SNR can make the cesium beam clock still have excellent performance. Therefore, lasers with narrow linewidth and low frequency noise are essential.

In order to optimize the frequency stability of optically pumped cesium beam clocks, we use frequency-stabilized Faraday laser \cite{doi:10.1063/1.3624696,chang2017faraday,chang2019faraday,Rotondaro:18,Tang:21,9953040} as the pumping and detection laser sources. Compared with traditional external-cavity semiconductor lasers (ECDLs), such as lasers frequency selected by interferences and gratings, the Faraday laser has the advantages of immunity to temperature and current variations of the laser diode, and the frequency of Faraday laser can be automatically corresponding to atomic transition lines as soon as turning on, since Faraday laser is frequency selected by atomic filter\cite{doi:10.1063/1.3624696,chang2017faraday,chang2019faraday}. Moreover, the modulation transfer spectroscopy (MTS) technique\cite{MTS1,MTS2,MTS3} is applied to lock the Faraday laser. Compared with other frequency stabilization methods, such as the saturation absorption spectroscopy (SAS)\cite{SAS1,SAS2,SAS3}, polarization spectroscopy (PS)\cite{PS1,PS2}, dichroic atomic vapor laser
lock (DAVLL)\cite{DAVLL1,DAVLL2}, MTS technique effectively suppress low-frequency noise, and it has the characteristics of high SNR and  background-free\cite{MTS1}.


In this work, we achieve a high-performance Faraday laser pumped cesium beam clock, which applies an 852 nm Faraday laser frequency-stabilized by MTS technique as pumping and detection lasers. We measure the performance of the frequency-stabilized Faraday laser using the optical heterodyne method, and the result shows that the fractional frequency stability of each system is $1.8\times{10}^{-12}$ at 1 s, The most probable linewidth of the heterodyne beating signal is 12.7 kHz when Faraday lasers are free-running, which means that of the single system is 9.0 kHz. The most probable linewidth is 3.5 kHz with Faraday lasers being frequency locked by MTS, thus that of the single system is 2.5 kHz. We use such Faraday laser with narrow linewidth and low frequency noise as the pumping and the detection lasers to build the compact optically pumped cesium beam atomic clock, leading to the SNR of the Ramsey fringe being optimized. Thus, the fractional frequency stability of the Faraday laser pumped cesium beam clock can achieve $7.9\times{10}^{-13}$ at 1 s and $1.3\times{10}^{-13}$ at 100 s, dropping to $1.4\times{10}^{-14}$ at 10000 s when the cesium oven temperature is 110°C. 

\section{Experiment}

\begin{figure}[htbp]
	\subfigure{
		{\label{1.a}}
		\begin{overpic}[width=0.48\textwidth]{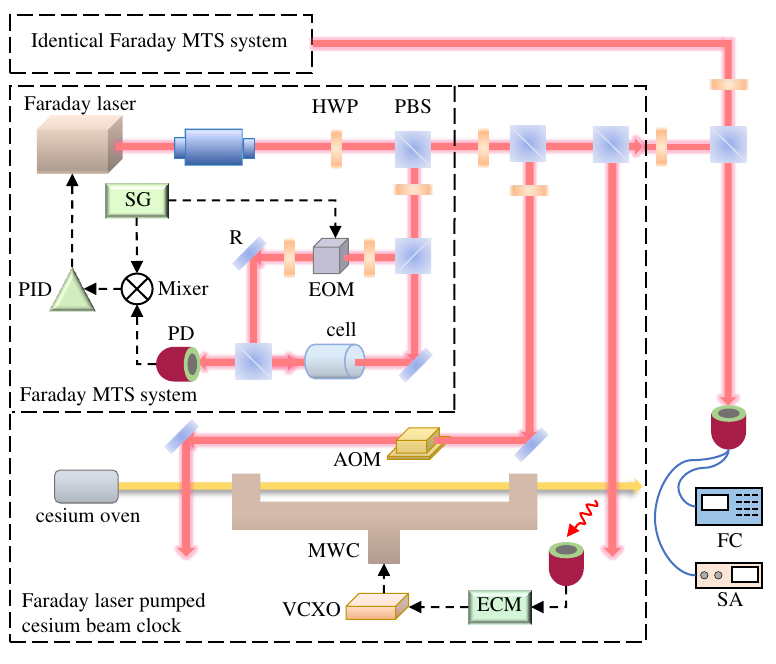}
			\put(-9,195){(a)}
		\end{overpic}
	}
	\linebreak
	\linebreak
	\subfigure{
		{\label{1.b}}
		\begin{overpic}[width=0.45\textwidth]{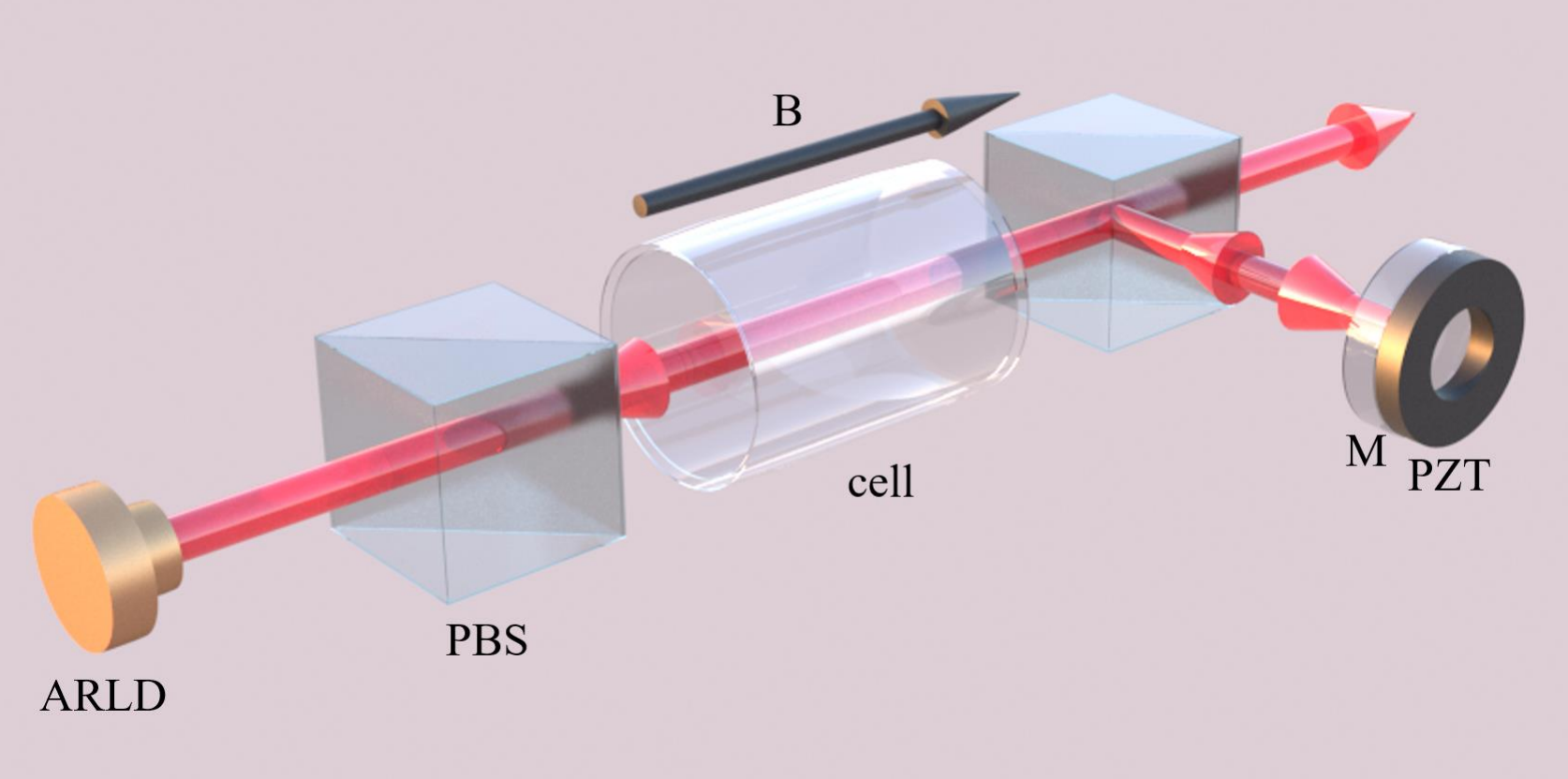}
			\put(-13,110){(b)}
		\end{overpic}
	}
	\linebreak
	\linebreak
	\subfigure{
		{\label{1.c}}
		\begin{overpic}[width=0.3\textwidth]{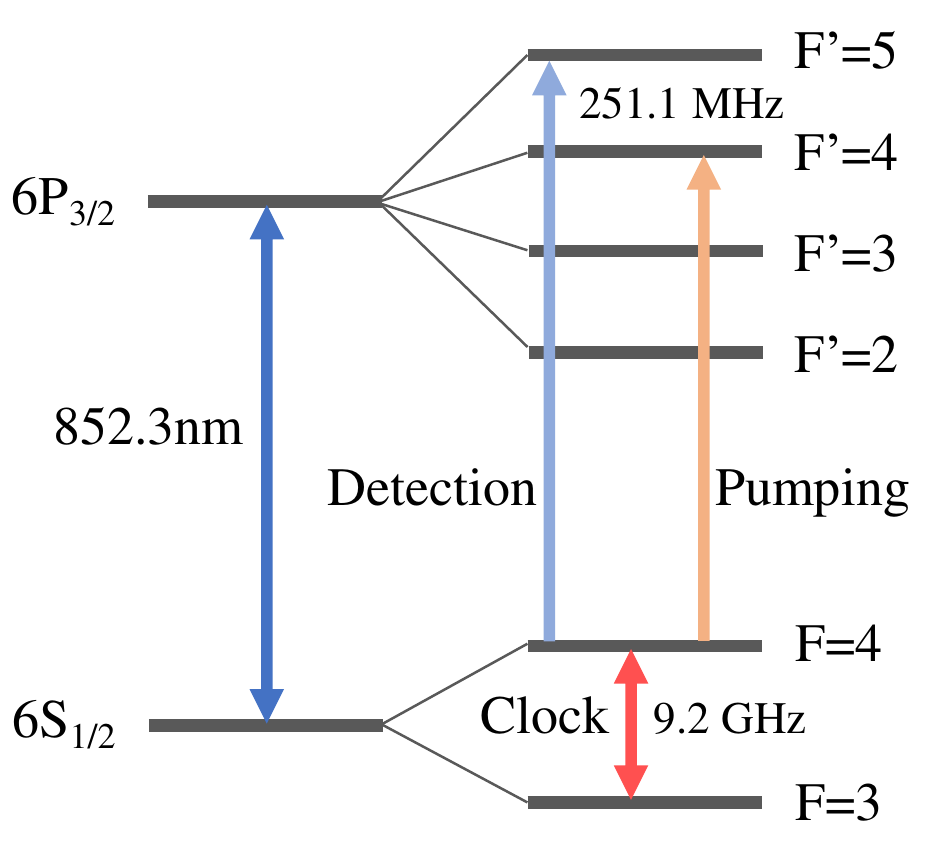}
			\put(-52,130){(c)}
		\end{overpic}
	}
\linebreak
	\caption{(a) Experimental setup. cell: cesium vapor cell of MTS, ISO: isolator, HWP: half wave plate, EOM: electro-optic modulator, PD: photoelectric detector, SG: signal generator, A: amplifier, PID: proportional integral derivative controller, ECM: electronic control module, VCXO: voltage-controlled crystal oscillator, MWC: microwave cavity, FC: frequency counter, SA: spectrum analyzer. (b) The schematic diagram of the Faraday laser. ARLD: anti-reflection coated laser diode, PBS: polarization beam splitter, cell: cesium vapor cell, M: high-reflectivity mirror, PZT: piezoelectric transducer, B: magnetic field. (c) Cesium relevant energy level diagram.}
\end{figure}

Figure \ref{1.a} shows the overall structure of the experiment. The Faraday laser is an integrated system with MTS technique for frequency stabilization, and all components are integrated onto a single aluminum base. The Faraday laser is beam-split and frequency-shifted to serve as the pumping and detection lasers for the cesium beam clock.

The structure of the Faraday laser is depicted in Fig. \ref{1.b}. We employ an anti-reflection coated laser diode (ARLD) as the gain medium. The anti-reflection coating superbly suppresses the impact of the internal cavity modes, which are formed between the front and rear surfaces of the ARLD, on the laser modes. Two PBSs, a cesium vapor cell, and permanent magnets compose the Faraday anomalous dispersion optical filter (FADOF) which is for frequency selection. The vapor cell is filled with cesium atoms and 10 torr argon buffer gas, with a diameter of 1.5 cm and a length of 3 cm. The cesium vapor cell is surrounded with a twisted copper heating wire driven by a heater temperature controller and enclosed by thermal insulation materials. The temperature of the cesium vapor cell is controlled at 66°C with a temperature fluctuation of 0.1°C.  The permanent magnets are placed around the cesium vapor cell to form a 900 G longitudinal magnetic field. Under the action of the magnetic field, the energy levels of the cesium atoms in the vapor cell undergo Zeeman splitting, which gives rise to the Faraday effect. The polarization direction of the linearly polarized light passing through the first PBS rotates as it propagates through the cesium vapor cell, and ultimately only the light whose polarization direction has rotated by $n\pi+\pi/2 (n=0,1,2\dots)$ can pass through the second PBS. Thus, only a very narrow frequency range of light corresponding to the spectral lines of the atoms can pass through the FADOF. Then the light is reflected back to the laser diode by a high-reflectivity mirror to form the external cavity feedback. Additionally, a piezoelectric transducer (PZT) is affixed to the high-reflectivity mirror, enabling the laser frequency to be tuned by adjusting the external cavity length. Due to the use of ARLD and the extremely narrow transmission spectrum of the FADOF, Faraday lasers are immune to the changes in current and temperature, a characteristic that distinguishes them from other types of ECDLs.



\begin{figure}
	\centering
	\includegraphics[width=0.5\textwidth]{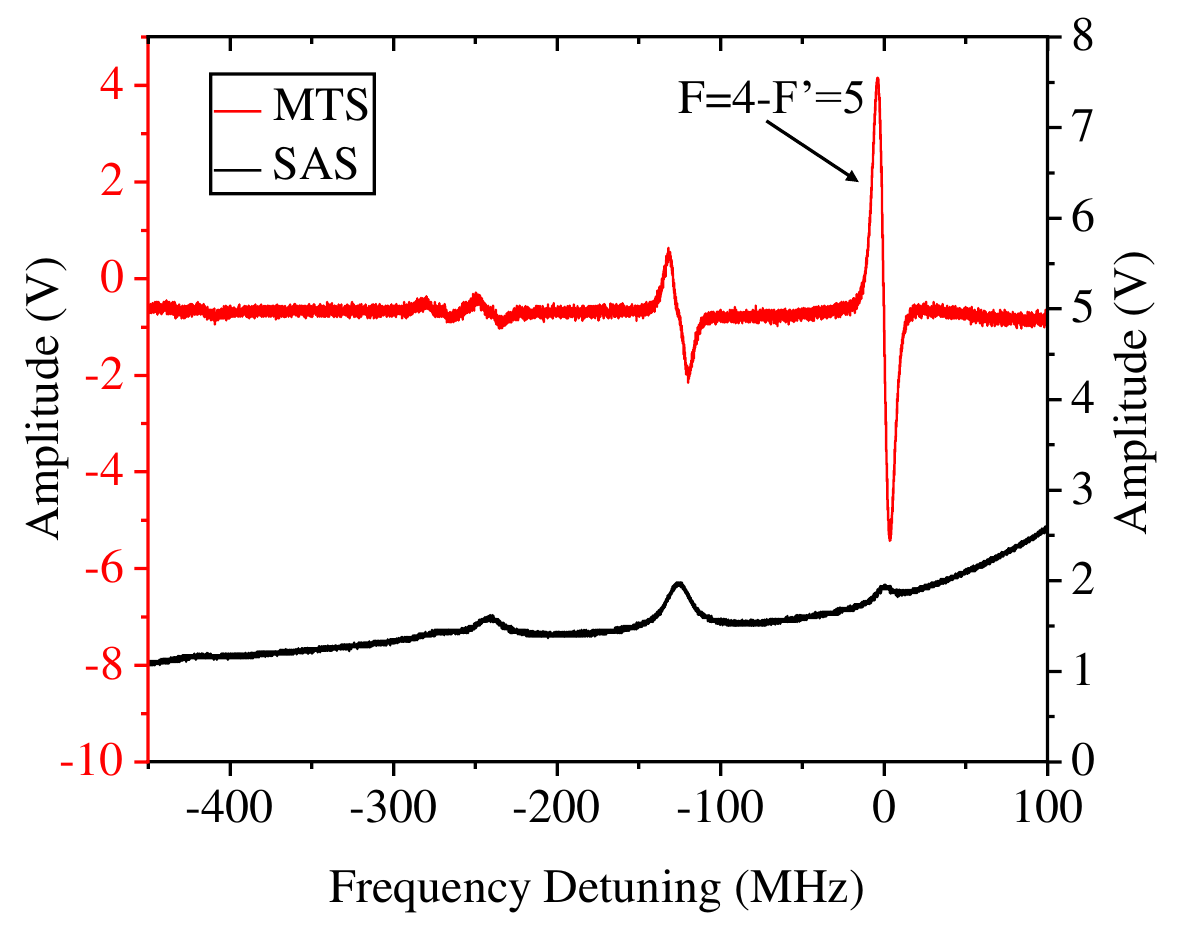}
	\caption{Modulation transfer spectroscopy (red line) and saturation absorption spectroscopy (black line) of the Faraday laser.}
	\label{MTS signal}
\end{figure}

In the MTS system, the frequency of the Faraday laser is locked to the cycling transition of $6s^2S_{1/2}|F=4\rangle - 6p^2P_{3/2}|F'=5\rangle$ by the MTS technique, as shown in Fig. \ref{1.c}. The Faraday laser beam is divided into pump laser and probe laser for MTS locking by a half wave plate (HWP) and a PBS. The pump laser is modulated by an electro-optic modulator (EOM) with a modulation frequency of 4.6 MHz. Then the modulation of the pump laser is transferred to the probe laser, which is received by the photoelectric detector (PD). The electrical signal of the PD is first amplified by an amplifier and then demodulated by a mixer with a demodulation frequency of 4.6 MHz to derive the MTS signal as shown in Fig. \ref{MTS signal}.  The MTS error signal is processed by a proportional integral derivative (PID) circuit (Vescent D2-125) to form a feedback signal, which is sent to the laser controller and the laser diode respectively for frequency stabilization. The former signal is processed by the laser controller and then fed back to the current and piezoelectric ceramic (PZT) of the Faraday laser. 

\begin{figure*}
	\centering
	\includegraphics[width=0.9\textwidth]{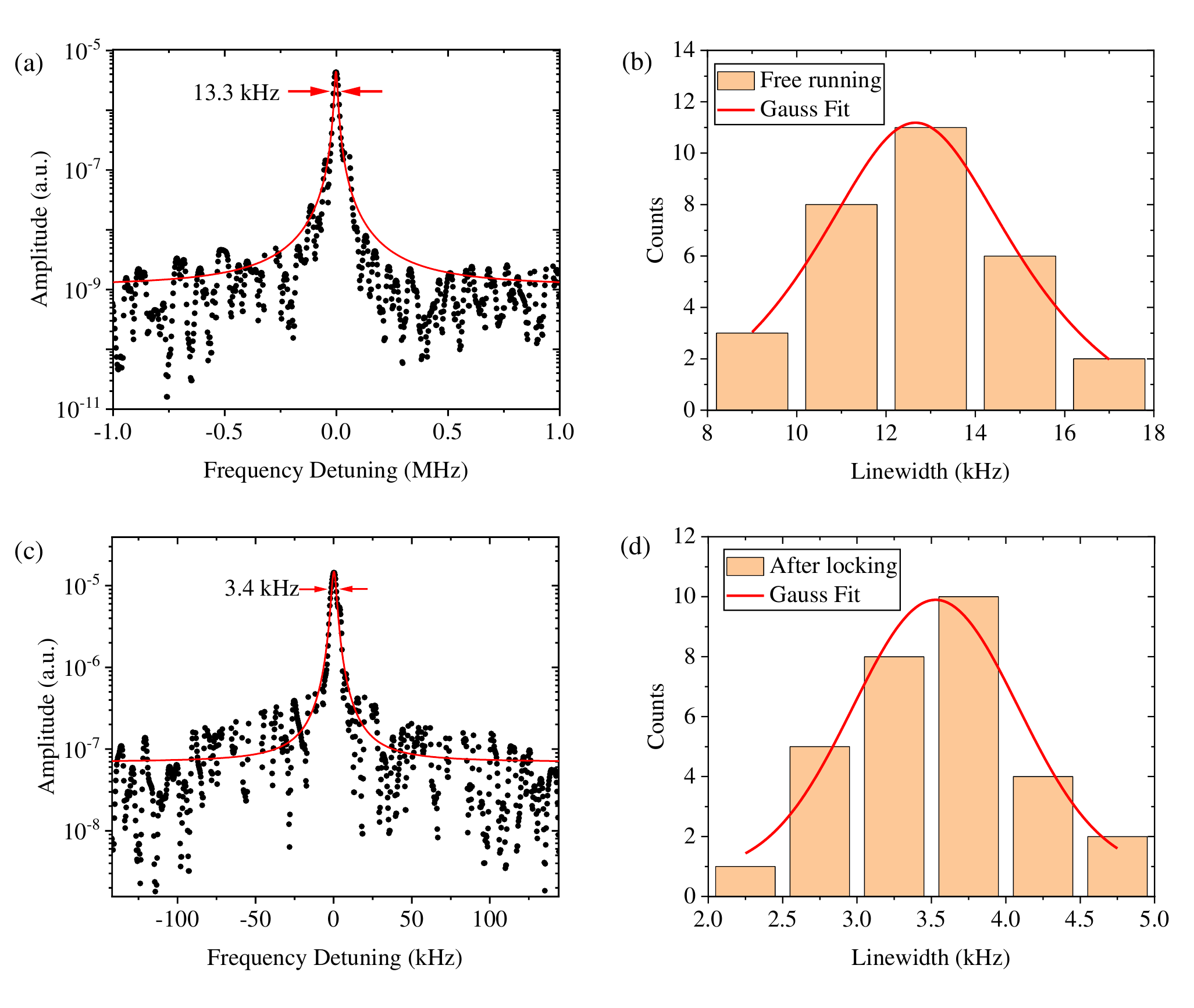}
	\caption{(a) The typical beating signal of two identical 852 nm Faraday lasers when they are free-running. (b) Histogram of repeated measurements of the fitted  beating linewidths when the two identical 852 nm Faraday lasers are free-running.  (c) The typical beating signal of two identical 852 nm MTS systems when they are locking. (d) Histogram of repeated measurements of the fitted beating linewidths when the two identical 852 nm MTS systems are locking. In figure (a) and (c), the black dots are the experimental results of the beating signal, and the red line is the Lorentz-fitting of the experimental results. In figure (b) and (d), the red line is the Gauss fitting of the histogram data.}
	\label{linewidth}
\end{figure*}

As shown in Fig. \ref{1.a}, we implement two identical 852 nm frequency-stabilized Faraday laser systems to realize the optical heterodyne measurement.
When both Faraday lasers are free-running, the signal derived by PD is sent into the spectrum analyzer, then we can obtain the fitted full width at half maximum (FWHM) of the heterodyne beating signal, as shown in Fig. \ref{linewidth}(a). 30 sets of data are measured to get the most probable linewidth of the free-running Faraday laser which is statistically calculated to be 12.7 kHz, so that the most available linewidth for a single laser is 9.0 kHz, as shown in Fig. \ref{linewidth}(b). When both laser systems are locked, the linewidth of the 852 nm MTS systems is obtained by the same method. The corresponding heterodyne beating signal obtained is shown in Fig. \ref{linewidth}(c). The most probable linewidth between the two frequency-stabilized Faraday lasers is 3.5 kHz, so the most allowable linewidth of a single laser is 2.5 kHz, as shown in Fig. \ref{linewidth}(d). The MTS technique greatly reduces the noise of the free-running Faraday laser. The narrow linewidth characteristic of the frequency-stabilized Faraday laser can significantly expand its application range.

\begin{figure}
	\centering
	\includegraphics[width=0.5\textwidth]{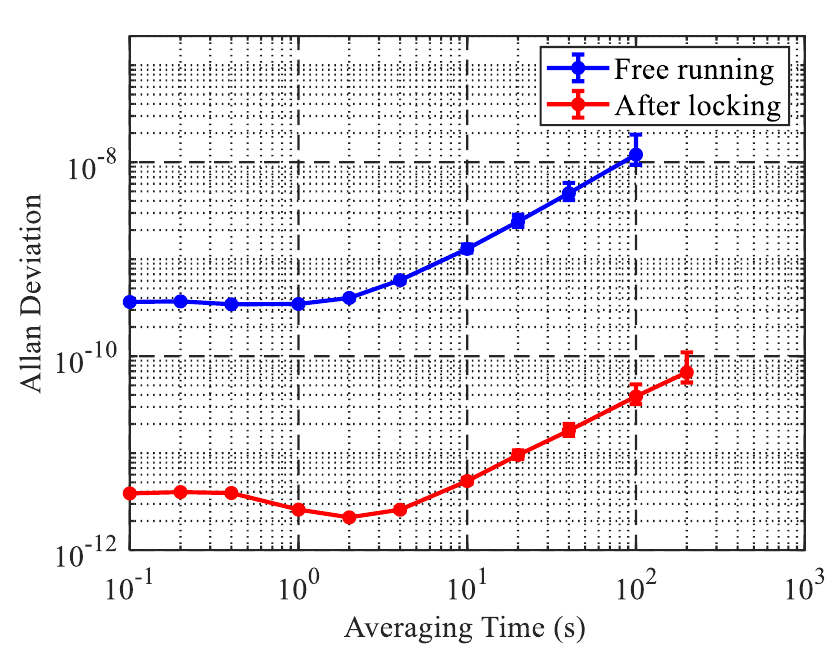}
	\caption{Frequency stability of the free-running Faraday laser and the frequency-stability Faraday laser measured by heterodyne measurement method. }
	\label{Allan}
\end{figure}

At the same time, the heterodyne beating signal from the PD is sent into a frequency counter (Keysight 53230A) to measure the frequency stability of the Faraday laser. The frequency stability of the free-running Faraday laser measured by the heterodyne method is $3.0 \times 10^{-10}$ at 1 s, as shown in Fig. \ref{Allan}. With both laser systems locked, the corresponding calculated frequency stability is shown in Fig. \ref{Allan}. The frequency stability of the heterodyne beating signal is  $2.6 \times 10^{-12}$ at 1 s, so the frequency stability of the single frequency-stabilized Faraday laser is $1.8 \times 10^{-12}$ at 1 s. In the MTS system, we use a double-layer vapor cell as the absolute frequency reference. Changes in the temperature of the atomic vapor cell bring about a drift in the frequency of the cesium transition lines, whereas the space between the two quartz layers of the double-layer vapor cell is pumped into a vacuum, and the double-layer vapor cell is therefore well insulated from the effect of ambient temperature fluctuations on the frequency of the cesium transition lines.


%

In the compact optically pumped cesium beam atomic clock, the pumping and the detection lasers are derived from the same Faraday laser frequency-stabilized on the cesium cycling transition. The output of the Faraday laser is split into two beams. One beam is used directly as the detection laser, while the other beam is frequency-shifted by 251.1 MHz using an acousto-optic modulator (AOM) and subsequently used as the pumping laser, which has a frequency corresponding to the $6s^2S_{1/2}|F=4\rangle - 6p^2P_{3/2}|F'=4\rangle$ transition line. The corresponding atomic transitions are schematically shown in Fig \ref{1.c}. Both the pumping and the detection lasers pass through the beam expanders before being injected into the cesium beam tube, enlarging the interaction area between the laser beams and the cesium beam by a factor of $\sim 10$. In the area where the cesium beam interacts with the pumping laser, a reflector is mounted on the back of the cesium beam tube to reflect the pumping laser to interact with the cesium beam again to improve the pumping efficiency.

\section{Results}

\begin{figure}
	\centering
	\includegraphics[width=0.5\textwidth]{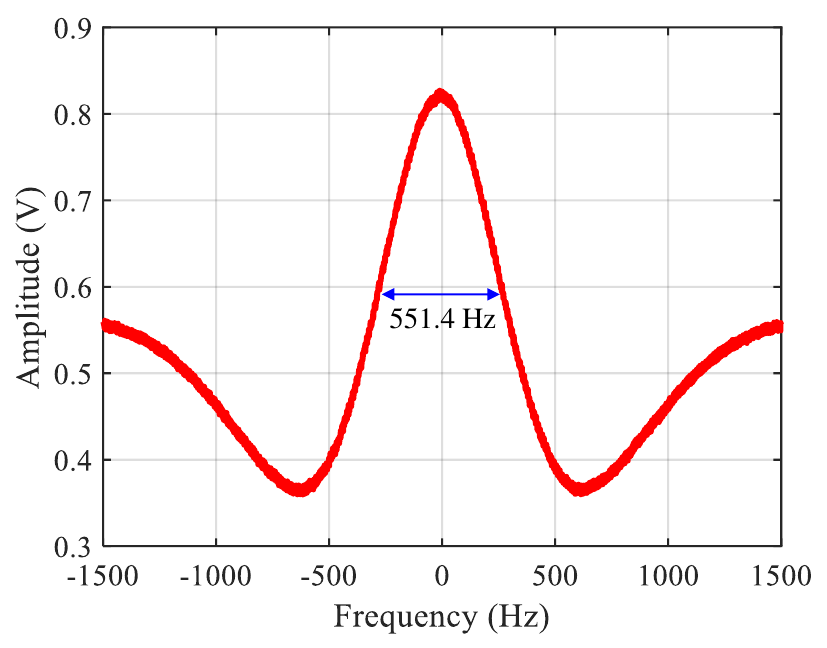}
	\caption{Ramsey fringe of the microwave clock transition which has a central frequency corresponding to the cesium transition $6s^2S_{1/2}|F=3, m_F=0\rangle - 6s^2S_{1/2}|F=4,m_F=0\rangle$. The linewidth of the Ramsey fringe is 551.4 Hz.}
	\label{Ramsey}
\end{figure}

A microwave signal synthesized by a 10 MHz voltage-controlled
crystal oscillator (VCXO) is acted on the microwave cavity and tuned with a center frequency of 9.192631770 GHz. The cesium oven temperature is set at 110 °C. The linewidth and the amplitude of the Ramsey fringe is 551.4 Hz and 461.85 mV, respectively, as shown in Fig. \ref{Ramsey}. The cesium beam clock operates at a modulation frequency of 137 Hz, and the noise of the Ramsey fringe signal is analyzed by a fast Fourier transform spectrum
analyzer (Stanford Research Systems, SR770). The noise at 137 Hz is $20.45 \  \rm{\mu V/\sqrt{Hz}}$. Thus, the SNR in 1 Hz bandwidth at 137 Hz is given by $\rm{SNR}=\textit{S/N}\approx 22600$, where $S$ represents the amplitude of the Ramsey fringe and $N$ represents the noise of the  Ramsey fringe at 137 Hz. The Ramsey fringe signal is modulated and demodulated to produce an error signal which is used to stabilize the output frequency of the VCXO by the servo circuits.

\begin{figure}
	\centering
	\includegraphics[width=0.5\textwidth]{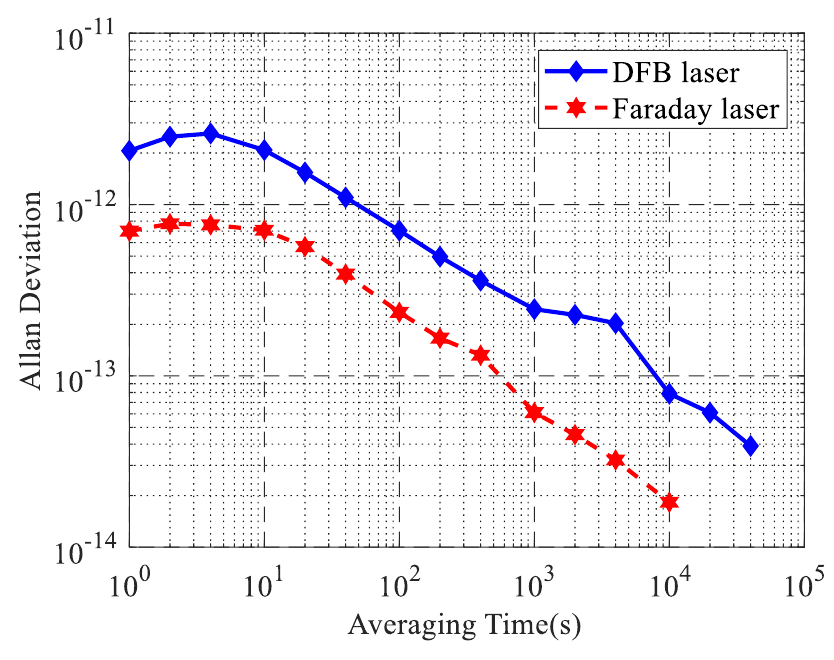}
	\caption{Frequency stability of the cesium beam clocks based on different types of lasers. The blue line shows the result in \cite{spaceon} used DFB laser. The red line shows the result of the Faraday laser pumped cesium beam clock.}
	\label{clock frequency stability_100}
\end{figure}

After locking the 10 MHz VCXO of the cesium beam clock, we compare the 10 MHz output of the VCXO with a hydrogen maser (VREMYA-CH, VCH-1003M, option L) by a high-performance phase noise and Allan deviation analyzer (Microsemi, 5120A). The frequency stability of the Faraday laser pumped cesium beam clock is two times better than that of the DFB laser-based cesium beam clock\cite{spaceon}, as shown in Fig. \ref{clock frequency stability_100}. In addition, the frequency stability of our cesium beam clock is also significantly better than that of the cesium beam clock based on the interferometer ECDL\cite{shanghaosen}, which is $3.7 \times 10^{-13}$ at 100 s. The three cesium beam clocks are different only in the categories of the lasers. That is, they use the same cesium beam tube and electronic system, and their cesium oven temperatures are all set at 100°C. Thus, the comparison results highlight the superiority of the Faraday laser, which has narrow linewidth and high frequency stability.

\begin{figure}
	\centering
	\includegraphics[width=0.5\textwidth]{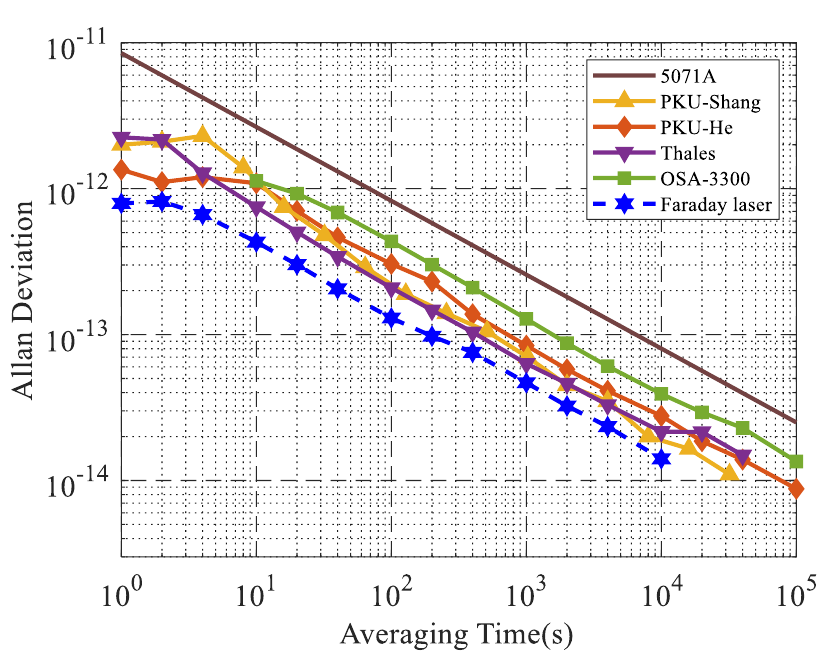}
		\caption{Frequency stability comparison of the cesium beam clocks. The brown line shows the frequency stability of the magnetic state selecting cesium beam clock called 5071A. The other lines show several frequency stabilities of optically pumped cesium beam clocks. The yellow line shows the result by Shang \textit{et al}.\cite{shanghaosen} The orange line shows the result by He \textit{et al}.\cite{chenxuzong} The purple line shows the result by Thales\cite{Galileo}. The green line shows the result of OSA-3300 by Oscilloquartz\cite{OSA-3300}. The blue line shows the result of the Faraday laser pumped cesium beam clock.}
	\label{clock frequency stability}
\end{figure}

We try different cesium oven temperatures, which would directly affect the SNR of the Ramsey fringe. As shown in Fig. \ref{clock frequency stability}, when the cesium oven temperature is increased to 110°C, the frequency stability of the cesium beam clock reaches $1.3 \times 10^{-12}/\sqrt{\tau}$, decreasing to $1.4\times{10}^{-14}$ at 10000 s. The frequency stability of our clock is better than that of any reported compact cesium beam atomic clocks to our knowledge, which is attributed to the narrow linewidth and the low frequency noise of the Faraday laser system resulting in a high SNR of the Ramsey fringe. Meanwhile, the frequency stability of our clock is $7.9 \times 10^{-13}$ at 1 s , which firstly reaches the $10^{-13}$ magnitude. When the cesium oven temperature is 130°C, the SNR of the Ramsey fringe in 1 Hz bandwidth at 137 Hz will reach  $39600 $, which can result in better frequency stability. Taking into account the lifetime of the cesium beam tube, the oven temperature is set at 110°C. The increase in signal intensity brought about by the increase in oven temperature does not necessarily lead directly to a higher SNR, which requires the support of a laser with low frequency noise\cite{10.1063/1.88159}.

In the future, we will apply auto-locking circuits to extend the measurement time of the cesium beam clock by automatically locking and re-locking the Faraday laser, which will fully demonstrate the excellent long-term performance of the cesium beam clock.

\section{Conclusion}
We implement a Faraday laser pumped cesium beam clock. Applying the Faraday laser to a cesium beam clock has several advantages. First, the Faraday laser is immune to the changes in current and temperature of the laser diode. In addition, because the Faraday laser uses atoms as frequency reference, the frequency of the Faraday laser immediately corresponds to the atomic spectra upon switching on and does not require manual tuning. Furthermore, the frequency-stabilized Faraday laser based on the MTS technique has narrow linewidth and high frequency stability. Measured by the optical heterodyne method, the linewidth of the frequency-stabilized Faraday laser is 2.5 kHz and the frequency stability of the frequency-stabilized Faraday laser is $1.8 \times 10^{-12}$ at 1 s due to the high SNR of the MTS signal. Thus, the application of the Faraday laser to cesium beam clock can enhance the SNR of Ramsey fringes, leading to the optimization of the cesium beam clock's performance. When the cesium oven temperature is 110°C, the cesium beam clock achieves an SNR in 1 Hz bandwidth  of $22600$, and the frequency stability of the cesium beam clock is $1.3\times{10}^{-12}/\sqrt{\tau}$, and  decreases to $1.4 \times 10^{-14}$ at 10000 s. When the cesium oven temperature is 130°C, the cesium beam clock achieves an SNR in 1 Hz bandwidth  of $39600$.  This Faraday laser pumped cesium beam clock has great potential and value in the fields of timekeeping, navigation, and communication. Meanwhile, the frequency-stabilized Faraday laser in this experiment can be applied in the field of precision measurement to further improve the performance of atomic clocks, atomic gravimeters, atomic magnetometers, and other applications. In our future work, we will aim to improve the atomic utilization efficiency of the cesium beam by hexapole magnetic system\cite{chenhaijun} and two-laser pumping scheme\cite{twolaser} and optimize the linewidth and noise of the Faraday laser to improve the performance of the cesium beam clock. 
\begin{acknowledgments}
	This study was funded by China Postdoctoral Science Foundation (BX2021020), Wenzhou Major Science \& Technology Innovation Key Project (ZG2020046), and Innovation Program for Quantum Science and Technology (2021ZD0303200).
\end{acknowledgments}
\bibliography{faraday}

\begin{thebibliography}{36}%
\makeatletter
\providecommand \@ifxundefined [1]{%
 \@ifx{#1\undefined}
}%
\providecommand \@ifnum [1]{%
 \ifnum #1\expandafter \@firstoftwo
 \else \expandafter \@secondoftwo
 \fi
}%
\providecommand \@ifx [1]{%
 \ifx #1\expandafter \@firstoftwo
 \else \expandafter \@secondoftwo
 \fi
}%
\providecommand \natexlab [1]{#1}%
\providecommand \enquote  [1]{``#1''}%
\providecommand \bibnamefont  [1]{#1}%
\providecommand \bibfnamefont [1]{#1}%
\providecommand \citenamefont [1]{#1}%
\providecommand \href@noop [0]{\@secondoftwo}%
\providecommand \href [0]{\begingroup \@sanitize@url \@href}%
\providecommand \@href[1]{\@@startlink{#1}\@@href}%
\providecommand \@@href[1]{\endgroup#1\@@endlink}%
\providecommand \@sanitize@url [0]{\catcode `\\12\catcode `\$12\catcode
  `\&12\catcode `\#12\catcode `\^12\catcode `\_12\catcode `\%12\relax}%
\providecommand \@@startlink[1]{}%
\providecommand \@@endlink[0]{}%
\providecommand \url  [0]{\begingroup\@sanitize@url \@url }%
\providecommand \@url [1]{\endgroup\@href {#1}{\urlprefix }}%
\providecommand \urlprefix  [0]{URL }%
\providecommand \Eprint [0]{\href }%
\providecommand \doibase [0]{https://doi.org/}%
\providecommand \selectlanguage [0]{\@gobble}%
\providecommand \bibinfo  [0]{\@secondoftwo}%
\providecommand \bibfield  [0]{\@secondoftwo}%
\providecommand \translation [1]{[#1]}%
\providecommand \BibitemOpen [0]{}%
\providecommand \bibitemStop [0]{}%
\providecommand \bibitemNoStop [0]{.\EOS\space}%
\providecommand \EOS [0]{\spacefactor3000\relax}%
\providecommand \BibitemShut  [1]{\csname bibitem#1\endcsname}%
\let\auto@bib@innerbib\@empty
\bibitem [{\citenamefont {Essen}\ and\ \citenamefont {Parry}(1955)}]{essen}%
  \BibitemOpen
  \bibfield  {author} {\bibinfo {author} {\bibfnamefont {L.}~\bibnamefont
  {Essen}}\ and\ \bibinfo {author} {\bibfnamefont {J.~V.~L.}\ \bibnamefont
  {Parry}},\ }\bibfield  {title} {\bibinfo {title} {An atomic standard of
  frequency and time interval: A cæsium resonator},\ }\href
  {https://doi.org/10.1038/176280a0} {\bibfield  {journal} {\bibinfo  {journal}
  {Nature}\ }\textbf {\bibinfo {volume} {176}},\ \bibinfo {pages} {280}
  (\bibinfo {year} {1955})}\BibitemShut {NoStop}%
\bibitem [{\citenamefont {Vanier}\ and\ \citenamefont
  {Audoin}(2005)}]{Vanier_2005}%
  \BibitemOpen
  \bibfield  {author} {\bibinfo {author} {\bibfnamefont {J.}~\bibnamefont
  {Vanier}}\ and\ \bibinfo {author} {\bibfnamefont {C.}~\bibnamefont
  {Audoin}},\ }\bibfield  {title} {\bibinfo {title} {The classical caesium beam
  frequency standard: fifty years later},\ }\href
  {https://doi.org/10.1088/0026-1394/42/3/S05} {\bibfield  {journal} {\bibinfo
  {journal} {Metrologia}\ }\textbf {\bibinfo {volume} {42}},\ \bibinfo {pages}
  {S31} (\bibinfo {year} {2005})}\BibitemShut {NoStop}%
\bibitem [{\citenamefont {Maleki}\ and\ \citenamefont
  {Prestage}(2005)}]{Maleki_2005}%
  \BibitemOpen
  \bibfield  {author} {\bibinfo {author} {\bibfnamefont {L.}~\bibnamefont
  {Maleki}}\ and\ \bibinfo {author} {\bibfnamefont {J.}~\bibnamefont
  {Prestage}},\ }\bibfield  {title} {\bibinfo {title} {Applications of clocks
  and frequency standards: from the routine to tests of fundamental models},\
  }\href {https://doi.org/10.1088/0026-1394/42/3/S15} {\bibfield  {journal}
  {\bibinfo  {journal} {Metrologia}\ }\textbf {\bibinfo {volume} {42}},\
  \bibinfo {pages} {S145} (\bibinfo {year} {2005})}\BibitemShut {NoStop}%
\bibitem [{\citenamefont {Essen}\ and\ \citenamefont
  {Parry}(1957)}]{essen1957caesium}%
  \BibitemOpen
  \bibfield  {author} {\bibinfo {author} {\bibfnamefont {L.}~\bibnamefont
  {Essen}}\ and\ \bibinfo {author} {\bibfnamefont {J.~V.~L.}\ \bibnamefont
  {Parry}},\ }\bibfield  {title} {\bibinfo {title} {The caesium resonator as a
  standard of frequency and time},\ }\href
  {https://doi.org/10.1098/rsta.1957.0010} {\bibfield  {journal} {\bibinfo
  {journal} {Philosophical Transactions of the Royal Society of London. Series
  A, Mathematical and Physical Sciences}\ }\textbf {\bibinfo {volume} {250}},\
  \bibinfo {pages} {45} (\bibinfo {year} {1957})}\BibitemShut {NoStop}%
\bibitem [{\citenamefont {Forman}(1985)}]{1457534}%
  \BibitemOpen
  \bibfield  {author} {\bibinfo {author} {\bibfnamefont {P.}~\bibnamefont
  {Forman}},\ }\bibfield  {title} {\bibinfo {title} {Atomichron®: The atomic
  clock from concept to commercial product},\ }\href
  {https://doi.org/10.1109/PROC.1985.13266} {\bibfield  {journal} {\bibinfo
  {journal} {Proceedings of the IEEE}\ }\textbf {\bibinfo {volume} {73}},\
  \bibinfo {pages} {1181} (\bibinfo {year} {1985})}\BibitemShut {NoStop}%
\bibitem [{\citenamefont {Audoin}(1992)}]{Audoin}%
  \BibitemOpen
  \bibfield  {author} {\bibinfo {author} {\bibfnamefont {C.}~\bibnamefont
  {Audoin}},\ }\bibfield  {title} {\bibinfo {title} {Caesium beam frequency
  standards: Classical and optically pumped},\ }\href
  {https://doi.org/10.1088/0026-1394/29/2/003} {\bibfield  {journal} {\bibinfo
  {journal} {Metrologia}\ }\textbf {\bibinfo {volume} {29}},\ \bibinfo {pages}
  {113} (\bibinfo {year} {1992})}\BibitemShut {NoStop}%
\bibitem [{\citenamefont {Chen}\ \emph
  {et~al.}(2022{\natexlab{a}})\citenamefont {Chen}, \citenamefont {Wang},
  \citenamefont {Guo}, \citenamefont {Yang}, \citenamefont {Ma}, \citenamefont
  {Huang},\ and\ \citenamefont {Liu}}]{510}%
  \BibitemOpen
  \bibfield  {author} {\bibinfo {author} {\bibfnamefont {J.}~\bibnamefont
  {Chen}}, \bibinfo {author} {\bibfnamefont {J.}~\bibnamefont {Wang}}, \bibinfo
  {author} {\bibfnamefont {L.}~\bibnamefont {Guo}}, \bibinfo {author}
  {\bibfnamefont {J.}~\bibnamefont {Yang}}, \bibinfo {author} {\bibfnamefont
  {P.}~\bibnamefont {Ma}}, \bibinfo {author} {\bibfnamefont {L.}~\bibnamefont
  {Huang}},\ and\ \bibinfo {author} {\bibfnamefont {Z.}~\bibnamefont {Liu}},\
  }\bibfield  {title} {\bibinfo {title} {Characteristics analysis of compact
  cesium atomic clock with magnetic state selection},\ }\href
  {https://doi.org/10.3389/fphy.2022.959343} {\bibfield  {journal} {\bibinfo
  {journal} {Front. Phys.}\ }\textbf {\bibinfo {volume} {10}},\ \bibinfo
  {pages} {959343} (\bibinfo {year} {2022}{\natexlab{a}})}\BibitemShut
  {NoStop}%
\bibitem [{\citenamefont {Lutwak}\ \emph {et~al.}(2001)\citenamefont {Lutwak},
  \citenamefont {Emmons}, \citenamefont {Garvey},\ and\ \citenamefont
  {Vlitas}}]{GPS_3}%
  \BibitemOpen
  \bibfield  {author} {\bibinfo {author} {\bibfnamefont {R.}~\bibnamefont
  {Lutwak}}, \bibinfo {author} {\bibfnamefont {D.}~\bibnamefont {Emmons}},
  \bibinfo {author} {\bibfnamefont {R.~M.}\ \bibnamefont {Garvey}},\ and\
  \bibinfo {author} {\bibfnamefont {P.}~\bibnamefont {Vlitas}},\ }\bibfield
  {title} {\bibinfo {title} {Optically {P}umped {C}esium-{B}eam {F}requency
  {S}tandard for {Gps}-\uppercase\expandafter{\romannumeral3}},\ }in\ \href
  {https://www.researchgate.net/publication/235052911_Optically_Pumped_Cesium-Beam_Frequency_Standard_for_GPS-III}
  {\emph {\bibinfo {booktitle} {Proceedings of the 33th Annual Precise Time and
  Time Interval Systems and Applications Meeting}}}\ (\bibinfo {year} {2001})\
  pp.\ \bibinfo {pages} {19--32}\BibitemShut {NoStop}%
\bibitem [{\citenamefont {Guérandel}\ \emph {et~al.}(2002)\citenamefont
  {Guérandel}, \citenamefont {Hermann}, \citenamefont {Barillet},
  \citenamefont {Cerez}, \citenamefont {Théobald},\ and\ \citenamefont
  {et~al.}}]{2002}%
  \BibitemOpen
  \bibfield  {author} {\bibinfo {author} {\bibfnamefont {S.}~\bibnamefont
  {Guérandel}}, \bibinfo {author} {\bibfnamefont {V.}~\bibnamefont {Hermann}},
  \bibinfo {author} {\bibfnamefont {R.}~\bibnamefont {Barillet}}, \bibinfo
  {author} {\bibfnamefont {P.}~\bibnamefont {Cerez}}, \bibinfo {author}
  {\bibfnamefont {G.}~\bibnamefont {Théobald}},\ and\ \bibinfo {author}
  {\bibnamefont {et~al.}},\ }\bibfield  {title} {\bibinfo {title} {Compact
  cesium beam frequency standard: improvement of the frequency stability
  towards the $10^{-12} \tau^{-1/2}$ level.},\ }in\ \href
  {https://hal.science/hal-03734412/} {\emph {\bibinfo {booktitle} {Proceedings
  of the 16th European Frequency and Time Forum}}}\ (\bibinfo {year}
  {2002})\BibitemShut {NoStop}%
\bibitem [{\citenamefont {Sallot}\ \emph {et~al.}(2003)\citenamefont {Sallot},
  \citenamefont {Baldy}, \citenamefont {Gin},\ and\ \citenamefont
  {Petit}}]{Galileo_2003}%
  \BibitemOpen
  \bibfield  {author} {\bibinfo {author} {\bibfnamefont {C.}~\bibnamefont
  {Sallot}}, \bibinfo {author} {\bibfnamefont {M.}~\bibnamefont {Baldy}},
  \bibinfo {author} {\bibfnamefont {D.}~\bibnamefont {Gin}},\ and\ \bibinfo
  {author} {\bibfnamefont {R.}~\bibnamefont {Petit}},\ }\bibfield  {title}
  {\bibinfo {title} {$3 \cdot 10^{-12} \cdot \tau^{-1/2}$ on industrial
  prototype optically pumped cesium beam frequency standard},\ }in\ \href
  {https://doi.org/10.1109/FREQ.2003.1274995} {\emph {\bibinfo {booktitle}
  {IEEE International Frequency Control Symposium and PDA Exhibition Jointly
  with the 17th European Frequency and Time Forum}}}\ (\bibinfo {year} {2003})\
  pp.\ \bibinfo {pages} {100--104}\BibitemShut {NoStop}%
\bibitem [{\citenamefont {Hermann}\ \emph {et~al.}(2006)\citenamefont
  {Hermann}, \citenamefont {Leger}, \citenamefont {Vian}, \citenamefont
  {Jarno},\ and\ \citenamefont {Gazard}}]{Thales_2006}%
  \BibitemOpen
  \bibfield  {author} {\bibinfo {author} {\bibfnamefont {V.}~\bibnamefont
  {Hermann}}, \bibinfo {author} {\bibfnamefont {B.}~\bibnamefont {Leger}},
  \bibinfo {author} {\bibfnamefont {C.}~\bibnamefont {Vian}}, \bibinfo {author}
  {\bibfnamefont {J.-F.}\ \bibnamefont {Jarno}},\ and\ \bibinfo {author}
  {\bibfnamefont {M.}~\bibnamefont {Gazard}},\ }\bibfield  {title} {\bibinfo
  {title} {Industrial development of an optically pumped {C}s beam frequency
  standard for high performance applications},\ }in\ \href
  {https://ieeexplore.ieee.org/abstract/document/6231019} {\emph {\bibinfo
  {booktitle} {Proceedings of the 20th European Frequency and Time Forum}}}\
  (\bibinfo {year} {2006})\ pp.\ \bibinfo {pages} {432--435}\BibitemShut
  {NoStop}%
\bibitem [{\citenamefont {Schmeissner}\ \emph {et~al.}(2017)\citenamefont
  {Schmeissner}, \citenamefont {Favard}, \citenamefont {Douahi}, \citenamefont
  {Perez}, \citenamefont {Mestre}, \citenamefont {Baldy}, \citenamefont
  {Romer}, \citenamefont {Chastellain}, \citenamefont {Coppoolse},
  \citenamefont {von Bandel}, \citenamefont {Garcia}, \citenamefont
  {Krakowski}, \citenamefont {Guérandel}, \citenamefont {Folco},\ and\
  \citenamefont {Konrad}}]{Galileo}%
  \BibitemOpen
  \bibfield  {author} {\bibinfo {author} {\bibfnamefont {R.}~\bibnamefont
  {Schmeissner}}, \bibinfo {author} {\bibfnamefont {P.}~\bibnamefont {Favard}},
  \bibinfo {author} {\bibfnamefont {A.}~\bibnamefont {Douahi}}, \bibinfo
  {author} {\bibfnamefont {P.}~\bibnamefont {Perez}}, \bibinfo {author}
  {\bibfnamefont {N.}~\bibnamefont {Mestre}}, \bibinfo {author} {\bibfnamefont
  {M.}~\bibnamefont {Baldy}}, \bibinfo {author} {\bibfnamefont
  {A.}~\bibnamefont {Romer}}, \bibinfo {author} {\bibfnamefont
  {F.}~\bibnamefont {Chastellain}}, \bibinfo {author} {\bibfnamefont {W.~W.}\
  \bibnamefont {Coppoolse}}, \bibinfo {author} {\bibfnamefont {N.}~\bibnamefont
  {von Bandel}}, \bibinfo {author} {\bibfnamefont {M.}~\bibnamefont {Garcia}},
  \bibinfo {author} {\bibfnamefont {M.}~\bibnamefont {Krakowski}}, \bibinfo
  {author} {\bibfnamefont {S.}~\bibnamefont {Guérandel}}, \bibinfo {author}
  {\bibfnamefont {Y.}~\bibnamefont {Folco}},\ and\ \bibinfo {author}
  {\bibfnamefont {W.}~\bibnamefont {Konrad}},\ }\bibfield  {title} {\bibinfo
  {title} {Optically pumped {C}s space clock development},\ }in\ \href
  {https://doi.org/10.1109/FCS.2017.8088825} {\emph {\bibinfo {booktitle} {2017
  Joint Conference of the European Frequency and Time Forum and IEEE
  International Frequency Control Symposium (EFTF/IFCS)}}}\ (\bibinfo {year}
  {2017})\ pp.\ \bibinfo {pages} {136--137}\BibitemShut {NoStop}%
\bibitem [{\citenamefont {Shang}\ \emph {et~al.}(2020)\citenamefont {Shang},
  \citenamefont {Zhang}, \citenamefont {Miao}, \citenamefont {Shi},
  \citenamefont {Pan}, \citenamefont {Zhao}, \citenamefont {Wei}, \citenamefont
  {Yang},\ and\ \citenamefont {Chen}}]{shanghaosen}%
  \BibitemOpen
  \bibfield  {author} {\bibinfo {author} {\bibfnamefont {H.}~\bibnamefont
  {Shang}}, \bibinfo {author} {\bibfnamefont {T.}~\bibnamefont {Zhang}},
  \bibinfo {author} {\bibfnamefont {J.}~\bibnamefont {Miao}}, \bibinfo {author}
  {\bibfnamefont {T.}~\bibnamefont {Shi}}, \bibinfo {author} {\bibfnamefont
  {D.}~\bibnamefont {Pan}}, \bibinfo {author} {\bibfnamefont {X.}~\bibnamefont
  {Zhao}}, \bibinfo {author} {\bibfnamefont {Q.}~\bibnamefont {Wei}}, \bibinfo
  {author} {\bibfnamefont {L.}~\bibnamefont {Yang}},\ and\ \bibinfo {author}
  {\bibfnamefont {J.}~\bibnamefont {Chen}},\ }\bibfield  {title} {\bibinfo
  {title} {Laser with $10^{-13}$ short-term instability for compact optically
  pumped cesium beam atomic clock},\ }\href {https://doi.org/10.1364/OE.381147}
  {\bibfield  {journal} {\bibinfo  {journal} {Opt. Express}\ }\textbf {\bibinfo
  {volume} {28}},\ \bibinfo {pages} {6868} (\bibinfo {year}
  {2020})}\BibitemShut {NoStop}%
\bibitem [{\citenamefont {Chen}\ \emph
  {et~al.}(2022{\natexlab{b}})\citenamefont {Chen}, \citenamefont {Yan},
  \citenamefont {Chen},\ and\ \citenamefont {Feng}}]{chenhaijun}%
  \BibitemOpen
  \bibfield  {author} {\bibinfo {author} {\bibfnamefont {H.}~\bibnamefont
  {Chen}}, \bibinfo {author} {\bibfnamefont {Y.}~\bibnamefont {Yan}}, \bibinfo
  {author} {\bibfnamefont {J.}~\bibnamefont {Chen}},\ and\ \bibinfo {author}
  {\bibfnamefont {J.}~\bibnamefont {Feng}},\ }\bibfield  {title} {\bibinfo
  {title} {Design of optically pumped cesium beam tube with hexapole magnetic
  system for longer lifetime and better {SNR}},\ }\href
  {https://doi.org/10.3389/fphy.2022.956719} {\bibfield  {journal} {\bibinfo
  {journal} {Front. Phys.}\ }\textbf {\bibinfo {volume} {10}},\ \bibinfo
  {pages} {956719} (\bibinfo {year} {2022}{\natexlab{b}})}\BibitemShut
  {NoStop}%
\bibitem [{\citenamefont {He}\ \emph {et~al.}(2022)\citenamefont {He},
  \citenamefont {Yuan}, \citenamefont {Chen}, \citenamefont {Fang},
  \citenamefont {Chen}, \citenamefont {Wang},\ and\ \citenamefont
  {Qi}}]{chenxuzong}%
  \BibitemOpen
  \bibfield  {author} {\bibinfo {author} {\bibfnamefont {X.}~\bibnamefont
  {He}}, \bibinfo {author} {\bibfnamefont {Z.}~\bibnamefont {Yuan}}, \bibinfo
  {author} {\bibfnamefont {J.}~\bibnamefont {Chen}}, \bibinfo {author}
  {\bibfnamefont {S.}~\bibnamefont {Fang}}, \bibinfo {author} {\bibfnamefont
  {X.}~\bibnamefont {Chen}}, \bibinfo {author} {\bibfnamefont {Q.}~\bibnamefont
  {Wang}},\ and\ \bibinfo {author} {\bibfnamefont {X.}~\bibnamefont {Qi}},\
  }\bibfield  {title} {\bibinfo {title} {Improvement of the short- and
  long-term stability of high performance portable optically pumped cesium beam
  atomic clock},\ }\href {https://doi.org/10.3389/fphy.2022.970030} {\bibfield
  {journal} {\bibinfo  {journal} {Front. Phys.}\ }\textbf {\bibinfo {volume}
  {10}},\ \bibinfo {pages} {970030} (\bibinfo {year} {2022})}\BibitemShut
  {NoStop}%
\bibitem [{\citenamefont {P.~Berthoud}\ and\ \citenamefont
  {Dolgovskiy.}(2023)}]{OSA-3300}%
  \BibitemOpen
  \bibfield  {author} {\bibinfo {author} {\bibfnamefont {F.~K.}\ \bibnamefont
  {P.~Berthoud}, \bibfnamefont {M.~Haldimann}}\ and\ \bibinfo {author}
  {\bibfnamefont {V.}~\bibnamefont {Dolgovskiy.}},\ }\bibfield  {title}
  {\bibinfo {title} {High performance industrial cesium beam clock},\ }in\
  \href {https://doi.org/10.13140/RG.2.2.10067.55840} {\emph {\bibinfo
  {booktitle} {Proceedings of the 4th IFSA Frequency \& Time Conference (IFTC
  2022)}}}\ (\bibinfo {year} {2023})\ pp.\ \bibinfo {pages} {5--6}\BibitemShut
  {NoStop}%
\bibitem [{\citenamefont {Dimarcq}\ \emph {et~al.}(1993)\citenamefont
  {Dimarcq}, \citenamefont {Giordano}, \citenamefont {Cerez},\ and\
  \citenamefont {Theobald}}]{278532}%
  \BibitemOpen
  \bibfield  {author} {\bibinfo {author} {\bibfnamefont {N.}~\bibnamefont
  {Dimarcq}}, \bibinfo {author} {\bibfnamefont {V.}~\bibnamefont {Giordano}},
  \bibinfo {author} {\bibfnamefont {P.}~\bibnamefont {Cerez}},\ and\ \bibinfo
  {author} {\bibfnamefont {G.}~\bibnamefont {Theobald}},\ }\bibfield  {title}
  {\bibinfo {title} {Analysis of the noise sources in an optically pumped
  cesium beam resonator},\ }\href {https://doi.org/10.1109/19.278532}
  {\bibfield  {journal} {\bibinfo  {journal} {IEEE Transactions on
  Instrumentation and Measurement}\ }\textbf {\bibinfo {volume} {42}},\
  \bibinfo {pages} {115} (\bibinfo {year} {1993})}\BibitemShut {NoStop}%
\bibitem [{\citenamefont {Miao}\ \emph {et~al.}(2011)\citenamefont {Miao},
  \citenamefont {Yin}, \citenamefont {Zhuang}, \citenamefont {Luo},
  \citenamefont {Dang}, \citenamefont {Chen},\ and\ \citenamefont
  {Guo}}]{doi:10.1063/1.3624696}%
  \BibitemOpen
  \bibfield  {author} {\bibinfo {author} {\bibfnamefont {X.}~\bibnamefont
  {Miao}}, \bibinfo {author} {\bibfnamefont {L.}~\bibnamefont {Yin}}, \bibinfo
  {author} {\bibfnamefont {W.}~\bibnamefont {Zhuang}}, \bibinfo {author}
  {\bibfnamefont {B.}~\bibnamefont {Luo}}, \bibinfo {author} {\bibfnamefont
  {A.}~\bibnamefont {Dang}}, \bibinfo {author} {\bibfnamefont {J.}~\bibnamefont
  {Chen}},\ and\ \bibinfo {author} {\bibfnamefont {H.}~\bibnamefont {Guo}},\
  }\bibfield  {title} {\bibinfo {title} {Note: Demonstration of an
  external-cavity diode laser system immune to current and temperature
  fluctuations},\ }\href {https://doi.org/10.1063/1.3624696} {\bibfield
  {journal} {\bibinfo  {journal} {Rev. Sci. Instrum.}\ }\textbf {\bibinfo
  {volume} {82}},\ \bibinfo {pages} {086106} (\bibinfo {year}
  {2011})}\BibitemShut {NoStop}%
\bibitem [{\citenamefont {Chang}\ \emph {et~al.}(2017)\citenamefont {Chang},
  \citenamefont {Peng}, \citenamefont {Zhang}, \citenamefont {Chen},
  \citenamefont {Luo}, \citenamefont {Chen},\ and\ \citenamefont
  {Guo}}]{chang2017faraday}%
  \BibitemOpen
  \bibfield  {author} {\bibinfo {author} {\bibfnamefont {P.}~\bibnamefont
  {Chang}}, \bibinfo {author} {\bibfnamefont {H.}~\bibnamefont {Peng}},
  \bibinfo {author} {\bibfnamefont {S.}~\bibnamefont {Zhang}}, \bibinfo
  {author} {\bibfnamefont {Z.}~\bibnamefont {Chen}}, \bibinfo {author}
  {\bibfnamefont {B.}~\bibnamefont {Luo}}, \bibinfo {author} {\bibfnamefont
  {J.}~\bibnamefont {Chen}},\ and\ \bibinfo {author} {\bibfnamefont
  {H.}~\bibnamefont {Guo}},\ }\bibfield  {title} {\bibinfo {title} {A {F}araday
  laser lasing on {R}b 1529 nm transition},\ }\href
  {https://doi.org/10.1038/s41598-017-09501-w} {\bibfield  {journal} {\bibinfo
  {journal} {Sci. Rep.}\ }\textbf {\bibinfo {volume} {7}},\ \bibinfo {pages}
  {1} (\bibinfo {year} {2017})}\BibitemShut {NoStop}%
\bibitem [{\citenamefont {Chang}\ \emph {et~al.}(2019)\citenamefont {Chang},
  \citenamefont {Chen}, \citenamefont {Shang}, \citenamefont {Guan},
  \citenamefont {Guo}, \citenamefont {Chen},\ and\ \citenamefont
  {Luo}}]{chang2019faraday}%
  \BibitemOpen
  \bibfield  {author} {\bibinfo {author} {\bibfnamefont {P.}~\bibnamefont
  {Chang}}, \bibinfo {author} {\bibfnamefont {Y.}~\bibnamefont {Chen}},
  \bibinfo {author} {\bibfnamefont {H.}~\bibnamefont {Shang}}, \bibinfo
  {author} {\bibfnamefont {X.}~\bibnamefont {Guan}}, \bibinfo {author}
  {\bibfnamefont {H.}~\bibnamefont {Guo}}, \bibinfo {author} {\bibfnamefont
  {J.}~\bibnamefont {Chen}},\ and\ \bibinfo {author} {\bibfnamefont
  {B.}~\bibnamefont {Luo}},\ }\bibfield  {title} {\bibinfo {title} {A {F}araday
  laser operating on {C}s 852 nm transition},\ }\href
  {https://doi.org/10.1007/s00340-019-7342-5} {\bibfield  {journal} {\bibinfo
  {journal} {Appl. Phys. B}\ }\textbf {\bibinfo {volume} {125}},\ \bibinfo
  {pages} {1} (\bibinfo {year} {2019})}\BibitemShut {NoStop}%
\bibitem [{\citenamefont {Rotondaro}\ \emph {et~al.}(2018)\citenamefont
  {Rotondaro}, \citenamefont {Zhdanov}, \citenamefont {Shaffer},\ and\
  \citenamefont {Knize}}]{Rotondaro:18}%
  \BibitemOpen
  \bibfield  {author} {\bibinfo {author} {\bibfnamefont {M.~D.}\ \bibnamefont
  {Rotondaro}}, \bibinfo {author} {\bibfnamefont {B.~V.}\ \bibnamefont
  {Zhdanov}}, \bibinfo {author} {\bibfnamefont {M.~K.}\ \bibnamefont
  {Shaffer}},\ and\ \bibinfo {author} {\bibfnamefont {R.}~\bibnamefont
  {Knize}},\ }\bibfield  {title} {\bibinfo {title} {Narrowband diode laser pump
  module for pumping alkali vapors},\ }\href
  {https://doi.org/10.1364/OE.26.009792} {\bibfield  {journal} {\bibinfo
  {journal} {Opt. Express}\ }\textbf {\bibinfo {volume} {26}},\ \bibinfo
  {pages} {9792} (\bibinfo {year} {2018})}\BibitemShut {NoStop}%
\bibitem [{\citenamefont {Tang}\ \emph {et~al.}(2021)\citenamefont {Tang},
  \citenamefont {Zhao}, \citenamefont {Wang}, \citenamefont {Li}, \citenamefont
  {Yang}, \citenamefont {Wang}, \citenamefont {Yang}, \citenamefont {Han},\
  and\ \citenamefont {Xu}}]{Tang:21}%
  \BibitemOpen
  \bibfield  {author} {\bibinfo {author} {\bibfnamefont {H.}~\bibnamefont
  {Tang}}, \bibinfo {author} {\bibfnamefont {H.}~\bibnamefont {Zhao}}, \bibinfo
  {author} {\bibfnamefont {R.}~\bibnamefont {Wang}}, \bibinfo {author}
  {\bibfnamefont {L.}~\bibnamefont {Li}}, \bibinfo {author} {\bibfnamefont
  {Z.}~\bibnamefont {Yang}}, \bibinfo {author} {\bibfnamefont {H.}~\bibnamefont
  {Wang}}, \bibinfo {author} {\bibfnamefont {W.}~\bibnamefont {Yang}}, \bibinfo
  {author} {\bibfnamefont {K.}~\bibnamefont {Han}},\ and\ \bibinfo {author}
  {\bibfnamefont {X.}~\bibnamefont {Xu}},\ }\bibfield  {title} {\bibinfo
  {title} {18{W} ultra-narrow diode laser absolutely locked to the {R}b {D}$_2$
  line},\ }\href {https://doi.org/10.1364/OE.442523} {\bibfield  {journal}
  {\bibinfo  {journal} {Opt. Express}\ }\textbf {\bibinfo {volume} {29}},\
  \bibinfo {pages} {38728} (\bibinfo {year} {2021})}\BibitemShut {NoStop}%
\bibitem [{\citenamefont {Shi}\ \emph {et~al.}(2022)\citenamefont {Shi},
  \citenamefont {Chang}, \citenamefont {Wang}, \citenamefont {Liu},
  \citenamefont {Shi},\ and\ \citenamefont {Chen}}]{9953040}%
  \BibitemOpen
  \bibfield  {author} {\bibinfo {author} {\bibfnamefont {H.}~\bibnamefont
  {Shi}}, \bibinfo {author} {\bibfnamefont {P.}~\bibnamefont {Chang}}, \bibinfo
  {author} {\bibfnamefont {Z.}~\bibnamefont {Wang}}, \bibinfo {author}
  {\bibfnamefont {Z.}~\bibnamefont {Liu}}, \bibinfo {author} {\bibfnamefont
  {T.}~\bibnamefont {Shi}},\ and\ \bibinfo {author} {\bibfnamefont
  {J.}~\bibnamefont {Chen}},\ }\bibfield  {title} {\bibinfo {title} {Frequency
  stabilization of a cesium {F}araday laser with a double-layer vapor cell as
  frequency reference},\ }\href {https://doi.org/10.1109/JPHOT.2022.3221494}
  {\bibfield  {journal} {\bibinfo  {journal} {IEEE Photonics Journal}\ }\textbf
  {\bibinfo {volume} {14}},\ \bibinfo {pages} {1} (\bibinfo {year}
  {2022})}\BibitemShut {NoStop}%
\bibitem [{\citenamefont {Raj}\ \emph {et~al.}(1980)\citenamefont {Raj},
  \citenamefont {Bloch}, \citenamefont {Snyder}, \citenamefont {Camy},\ and\
  \citenamefont {Ducloy}}]{MTS1}%
  \BibitemOpen
  \bibfield  {author} {\bibinfo {author} {\bibfnamefont {R.~K.}\ \bibnamefont
  {Raj}}, \bibinfo {author} {\bibfnamefont {D.}~\bibnamefont {Bloch}}, \bibinfo
  {author} {\bibfnamefont {J.~J.}\ \bibnamefont {Snyder}}, \bibinfo {author}
  {\bibfnamefont {G.}~\bibnamefont {Camy}},\ and\ \bibinfo {author}
  {\bibfnamefont {M.}~\bibnamefont {Ducloy}},\ }\bibfield  {title} {\bibinfo
  {title} {High-frequency optically heterodyned saturation spectroscopy via
  resonant degenerate four-wave mixing},\ }\href
  {https://doi.org/10.1103/PhysRevLett.44.1251} {\bibfield  {journal} {\bibinfo
   {journal} {Phys. Rev. Lett.}\ }\textbf {\bibinfo {volume} {44}},\ \bibinfo
  {pages} {1251} (\bibinfo {year} {1980})}\BibitemShut {NoStop}%
\bibitem [{\citenamefont {Ye}\ \emph {et~al.}(2001)\citenamefont {Ye},
  \citenamefont {Ma},\ and\ \citenamefont {Hall}}]{MTS2}%
  \BibitemOpen
  \bibfield  {author} {\bibinfo {author} {\bibfnamefont {J.}~\bibnamefont
  {Ye}}, \bibinfo {author} {\bibfnamefont {L.~S.}\ \bibnamefont {Ma}},\ and\
  \bibinfo {author} {\bibfnamefont {J.~L.}\ \bibnamefont {Hall}},\ }\bibfield
  {title} {\bibinfo {title} {Molecular iodine clock},\ }\href
  {https://doi.org/10.1103/PhysRevLett.87.270801} {\bibfield  {journal}
  {\bibinfo  {journal} {Phys. Rev. Lett.}\ }\textbf {\bibinfo {volume} {87}},\
  \bibinfo {pages} {270801} (\bibinfo {year} {2001})}\BibitemShut {NoStop}%
\bibitem [{\citenamefont {D\"oringshoff}\ \emph {et~al.}(2019)\citenamefont
  {D\"oringshoff}, \citenamefont {Gutsch}, \citenamefont {Schkolnik},
  \citenamefont {K\"urbis}, \citenamefont {Oswald}, \citenamefont {Pr\"obster},
  \citenamefont {Kovalchuk}, \citenamefont {Bawamia}, \citenamefont {Smol},
  \citenamefont {Schuldt}, \citenamefont {Lezius}, \citenamefont {Holzwarth},
  \citenamefont {Wicht}, \citenamefont {Braxmaier}, \citenamefont {Krutzik},\
  and\ \citenamefont {Peters}}]{MTS3}%
  \BibitemOpen
  \bibfield  {author} {\bibinfo {author} {\bibfnamefont {K.}~\bibnamefont
  {D\"oringshoff}}, \bibinfo {author} {\bibfnamefont {F.~B.}\ \bibnamefont
  {Gutsch}}, \bibinfo {author} {\bibfnamefont {V.}~\bibnamefont {Schkolnik}},
  \bibinfo {author} {\bibfnamefont {C.}~\bibnamefont {K\"urbis}}, \bibinfo
  {author} {\bibfnamefont {M.}~\bibnamefont {Oswald}}, \bibinfo {author}
  {\bibfnamefont {B.}~\bibnamefont {Pr\"obster}}, \bibinfo {author}
  {\bibfnamefont {E.~V.}\ \bibnamefont {Kovalchuk}}, \bibinfo {author}
  {\bibfnamefont {A.}~\bibnamefont {Bawamia}}, \bibinfo {author} {\bibfnamefont
  {R.}~\bibnamefont {Smol}}, \bibinfo {author} {\bibfnamefont {T.}~\bibnamefont
  {Schuldt}}, \bibinfo {author} {\bibfnamefont {M.}~\bibnamefont {Lezius}},
  \bibinfo {author} {\bibfnamefont {R.}~\bibnamefont {Holzwarth}}, \bibinfo
  {author} {\bibfnamefont {A.}~\bibnamefont {Wicht}}, \bibinfo {author}
  {\bibfnamefont {C.}~\bibnamefont {Braxmaier}}, \bibinfo {author}
  {\bibfnamefont {M.}~\bibnamefont {Krutzik}},\ and\ \bibinfo {author}
  {\bibfnamefont {A.}~\bibnamefont {Peters}},\ }\bibfield  {title} {\bibinfo
  {title} {Iodine frequency reference on a sounding rocket},\ }\href
  {https://doi.org/10.1103/PhysRevApplied.11.054068} {\bibfield  {journal}
  {\bibinfo  {journal} {Phys. Rev. Appl.}\ }\textbf {\bibinfo {volume} {11}},\
  \bibinfo {pages} {054068} (\bibinfo {year} {2019})}\BibitemShut {NoStop}%
\bibitem [{\citenamefont {Cuneo}\ \emph {et~al.}(1994)\citenamefont {Cuneo},
  \citenamefont {Maki},\ and\ \citenamefont {McIntyre}}]{SAS1}%
  \BibitemOpen
  \bibfield  {author} {\bibinfo {author} {\bibfnamefont {C.~J.}\ \bibnamefont
  {Cuneo}}, \bibinfo {author} {\bibfnamefont {J.~J.}\ \bibnamefont {Maki}},\
  and\ \bibinfo {author} {\bibfnamefont {D.~H.}\ \bibnamefont {McIntyre}},\
  }\bibfield  {title} {\bibinfo {title} {Optically stabilized diode laser using
  high‐contrast saturated absorption},\ }\href
  {https://doi.org/10.1063/1.111498} {\bibfield  {journal} {\bibinfo  {journal}
  {Appl. Phys. Lett.}\ }\textbf {\bibinfo {volume} {64}},\ \bibinfo {pages}
  {2625} (\bibinfo {year} {1994})}\BibitemShut {NoStop}%
\bibitem [{\citenamefont {Banerjee}\ and\ \citenamefont
  {Natarajan}(2003)}]{SAS2}%
  \BibitemOpen
  \bibfield  {author} {\bibinfo {author} {\bibfnamefont {A.}~\bibnamefont
  {Banerjee}}\ and\ \bibinfo {author} {\bibfnamefont {V.}~\bibnamefont
  {Natarajan}},\ }\bibfield  {title} {\bibinfo {title} {Saturated-absorption
  spectroscopy: eliminating crossover resonances by use of copropagating
  beams},\ }\href {https://doi.org/10.1364/OL.28.001912} {\bibfield  {journal}
  {\bibinfo  {journal} {Opt. Lett.}\ }\textbf {\bibinfo {volume} {28}},\
  \bibinfo {pages} {1912} (\bibinfo {year} {2003})}\BibitemShut {NoStop}%
\bibitem [{\citenamefont {Hummon}\ \emph {et~al.}(2018)\citenamefont {Hummon},
  \citenamefont {Kang}, \citenamefont {Bopp}, \citenamefont {Li}, \citenamefont
  {Westly}, \citenamefont {Kim}, \citenamefont {Fredrick}, \citenamefont
  {Diddams}, \citenamefont {Srinivasan}, \citenamefont {Aksyuk},\ and\
  \citenamefont {Kitching}}]{SAS3}%
  \BibitemOpen
  \bibfield  {author} {\bibinfo {author} {\bibfnamefont {M.~T.}\ \bibnamefont
  {Hummon}}, \bibinfo {author} {\bibfnamefont {S.}~\bibnamefont {Kang}},
  \bibinfo {author} {\bibfnamefont {D.~G.}\ \bibnamefont {Bopp}}, \bibinfo
  {author} {\bibfnamefont {Q.}~\bibnamefont {Li}}, \bibinfo {author}
  {\bibfnamefont {D.~A.}\ \bibnamefont {Westly}}, \bibinfo {author}
  {\bibfnamefont {S.}~\bibnamefont {Kim}}, \bibinfo {author} {\bibfnamefont
  {C.}~\bibnamefont {Fredrick}}, \bibinfo {author} {\bibfnamefont {S.~A.}\
  \bibnamefont {Diddams}}, \bibinfo {author} {\bibfnamefont {K.}~\bibnamefont
  {Srinivasan}}, \bibinfo {author} {\bibfnamefont {V.}~\bibnamefont {Aksyuk}},\
  and\ \bibinfo {author} {\bibfnamefont {J.~E.}\ \bibnamefont {Kitching}},\
  }\bibfield  {title} {\bibinfo {title} {Photonic chip for laser stabilization
  to an atomic vapor with $10^{-11}$ instability},\ }\href
  {https://doi.org/10.1364/OPTICA.5.000443} {\bibfield  {journal} {\bibinfo
  {journal} {Optica}\ }\textbf {\bibinfo {volume} {5}},\ \bibinfo {pages} {443}
  (\bibinfo {year} {2018})}\BibitemShut {NoStop}%
\bibitem [{\citenamefont {Hansch}\ and\ \citenamefont {Couillaud}(1980)}]{PS1}%
  \BibitemOpen
  \bibfield  {author} {\bibinfo {author} {\bibfnamefont {T.}~\bibnamefont
  {Hansch}}\ and\ \bibinfo {author} {\bibfnamefont {B.}~\bibnamefont
  {Couillaud}},\ }\bibfield  {title} {\bibinfo {title} {Laser frequency
  stabilization by polarization spectroscopy of a reflecting reference
  cavity},\ }\href
  {https://doi.org/https://doi.org/10.1016/0030-4018(80)90069-3} {\bibfield
  {journal} {\bibinfo  {journal} {Opt. Commun.}\ }\textbf {\bibinfo {volume}
  {35}},\ \bibinfo {pages} {441} (\bibinfo {year} {1980})}\BibitemShut
  {NoStop}%
\bibitem [{\citenamefont {Noh}(2012)}]{PS2}%
  \BibitemOpen
  \bibfield  {author} {\bibinfo {author} {\bibfnamefont {H.-R.}\ \bibnamefont
  {Noh}},\ }\bibfield  {title} {\bibinfo {title} {Lineshapes in two-color
  polarization spectroscopy for cesium},\ }\href
  {https://doi.org/10.1364/OE.20.021784} {\bibfield  {journal} {\bibinfo
  {journal} {Opt. Express}\ }\textbf {\bibinfo {volume} {20}},\ \bibinfo
  {pages} {21784} (\bibinfo {year} {2012})}\BibitemShut {NoStop}%
\bibitem [{\citenamefont {Su}\ \emph {et~al.}(2014)\citenamefont {Su},
  \citenamefont {Meng}, \citenamefont {Ji}, \citenamefont {Yuan}, \citenamefont
  {Zhao}, \citenamefont {Xiao},\ and\ \citenamefont {Jia}}]{DAVLL1}%
  \BibitemOpen
  \bibfield  {author} {\bibinfo {author} {\bibfnamefont {D.-Q.}\ \bibnamefont
  {Su}}, \bibinfo {author} {\bibfnamefont {T.-F.}\ \bibnamefont {Meng}},
  \bibinfo {author} {\bibfnamefont {Z.-H.}\ \bibnamefont {Ji}}, \bibinfo
  {author} {\bibfnamefont {J.-P.}\ \bibnamefont {Yuan}}, \bibinfo {author}
  {\bibfnamefont {Y.-T.}\ \bibnamefont {Zhao}}, \bibinfo {author}
  {\bibfnamefont {L.-T.}\ \bibnamefont {Xiao}},\ and\ \bibinfo {author}
  {\bibfnamefont {S.-T.}\ \bibnamefont {Jia}},\ }\bibfield  {title} {\bibinfo
  {title} {Application of sub-doppler {DAVLL} to laser frequency stabilization
  in atomic cesium},\ }\href {https://doi.org/10.1364/AO.53.007011} {\bibfield
  {journal} {\bibinfo  {journal} {Appl. Opt.}\ }\textbf {\bibinfo {volume}
  {53}},\ \bibinfo {pages} {7011} (\bibinfo {year} {2014})}\BibitemShut
  {NoStop}%
\bibitem [{\citenamefont {Millett-Sikking}\ \emph {et~al.}(2006)\citenamefont
  {Millett-Sikking}, \citenamefont {Hughes}, \citenamefont {Tierney},\ and\
  \citenamefont {Cornish}}]{DAVLL2}%
  \BibitemOpen
  \bibfield  {author} {\bibinfo {author} {\bibfnamefont {A.}~\bibnamefont
  {Millett-Sikking}}, \bibinfo {author} {\bibfnamefont {I.~G.}\ \bibnamefont
  {Hughes}}, \bibinfo {author} {\bibfnamefont {P.}~\bibnamefont {Tierney}},\
  and\ \bibinfo {author} {\bibfnamefont {S.~L.}\ \bibnamefont {Cornish}},\
  }\bibfield  {title} {\bibinfo {title} {{DAVLL} lineshapes in atomic
  rubidium},\ }\href {https://doi.org/10.1088/0953-4075/40/1/017} {\bibfield
  {journal} {\bibinfo  {journal} {Journal of Physics B: Atomic, Molecular and
  Optical Physics}\ }\textbf {\bibinfo {volume} {40}},\ \bibinfo {pages} {187}
  (\bibinfo {year} {2006})}\BibitemShut {NoStop}%
\bibitem [{\citenamefont {Cao}\ \emph {et~al.}(2017)\citenamefont {Cao},
  \citenamefont {Zhao}, \citenamefont {Xie}, \citenamefont {Wei}, \citenamefont
  {Yang}, \citenamefont {Chen},\ and\ \citenamefont {Zhang}}]{spaceon}%
  \BibitemOpen
  \bibfield  {author} {\bibinfo {author} {\bibfnamefont {Y.}~\bibnamefont
  {Cao}}, \bibinfo {author} {\bibfnamefont {X.}~\bibnamefont {Zhao}}, \bibinfo
  {author} {\bibfnamefont {W.}~\bibnamefont {Xie}}, \bibinfo {author}
  {\bibfnamefont {Q.}~\bibnamefont {Wei}}, \bibinfo {author} {\bibfnamefont
  {L.}~\bibnamefont {Yang}}, \bibinfo {author} {\bibfnamefont {H.}~\bibnamefont
  {Chen}},\ and\ \bibinfo {author} {\bibfnamefont {S.}~\bibnamefont {Zhang}},\
  }\bibfield  {title} {\bibinfo {title} {A merchandized optically pumped cesium
  atomic clock},\ }in\ \href {https://doi.org/10.1109/FCS.2017.8088976} {\emph
  {\bibinfo {booktitle} {2017 Joint Conference of the European Frequency and
  Time Forum and IEEE International Frequency Control Symposium (EFTF/IFCS)}}}\
  (\bibinfo {year} {2017})\ pp.\ \bibinfo {pages} {618--621}\BibitemShut
  {NoStop}%
\bibitem [{\citenamefont {Gerritsen}\ and\ \citenamefont
  {Nienhuis}(1975)}]{10.1063/1.88159}%
  \BibitemOpen
  \bibfield  {author} {\bibinfo {author} {\bibfnamefont {H.~J.}\ \bibnamefont
  {Gerritsen}}\ and\ \bibinfo {author} {\bibfnamefont {G.}~\bibnamefont
  {Nienhuis}},\ }\bibfield  {title} {\bibinfo {title} {{Multidirectional
  Doppler pumping: A new method to prepare an atomic beam having a large
  fraction of excited atoms}},\ }\href {https://doi.org/10.1063/1.88159}
  {\bibfield  {journal} {\bibinfo  {journal} {Appl. Phys. Lett.}\ }\textbf
  {\bibinfo {volume} {26}},\ \bibinfo {pages} {347} (\bibinfo {year}
  {1975})}\BibitemShut {NoStop}%
\bibitem [{\citenamefont {de~Clercq}\ \emph {et~al.}(1984)\citenamefont
  {de~Clercq}, \citenamefont {de~Labachellerie}, \citenamefont {Avila},
  \citenamefont {Cerez},\ and\ \citenamefont {Tetu}}]{twolaser}%
  \BibitemOpen
  \bibfield  {author} {\bibinfo {author} {\bibfnamefont {E.}~\bibnamefont
  {de~Clercq}}, \bibinfo {author} {\bibfnamefont {M.}~\bibnamefont
  {de~Labachellerie}}, \bibinfo {author} {\bibfnamefont {G.}~\bibnamefont
  {Avila}}, \bibinfo {author} {\bibfnamefont {P.}~\bibnamefont {Cerez}},\ and\
  \bibinfo {author} {\bibfnamefont {M.}~\bibnamefont {Tetu}},\ }\bibfield
  {title} {\bibinfo {title} {Laser diode optically pumped caesium beam},\
  }\href {https://doi.org/10.1051/jphys:01984004502023900} {\bibfield
  {journal} {\bibinfo  {journal} {J. de Phys.}\ }\textbf {\bibinfo {volume}
  {45}},\ \bibinfo {pages} {239} (\bibinfo {year} {1984})}\BibitemShut
  {NoStop}%
\end{thebibliography}%

\end{document}